\documentstyle{article}

\begin{document}

\baselineskip 0.6cm

\newcommand{\Bfp}{\mbox{\boldmath $p$}}
\newcommand{\Bfx}{\mbox{\boldmath $x$}}
\newcommand{\BfA}{\mbox{\boldmath $A$}}
\newcommand{\BfX}{\mbox{\boldmath $X$}}
\newcommand{\BfP}{\mbox{\boldmath $P$}}
\newcommand{\BfJ}{\mbox{\boldmath $J$}}
\newcommand{\Bfv}{\mbox{\boldmath $v$}}
\newcommand{\BfV}{\mbox{\boldmath $V$}}
\newcommand{\Bfe}{\mbox{\boldmath $e$}}
\newcommand{\BfK}{\mbox{\boldmath $K$}}
\newcommand{\BfR}{\mbox{\boldmath $R$}}
\newcommand{\sumtwo}[2]
   { \!\! \sum_{\begin{array}{c} {\scriptstyle #1} \\ {\scriptstyle #2}
   \end{array} } \!\! }
\newcommand{\sumtwoo}[2]
   { \!\! \sum^{N}_{\begin{array}{c} {\scriptstyle #1} \\ {\scriptstyle #2}
   \end{array} } \!\! }
\newcommand{\sumthree}[3]
   { \!\!\!\!\!\!\!\!\!\!\!\!\!\sum_{\begin{array}{c} {\scriptstyle #1} \\
   \left(\begin{array}{c} {\scriptstyle #2} \\
   {\scriptstyle #3} \end{array}\right) \end{array} }
   \!\!\!\!\!\!\!\!\!\!\!\!\!\!\! }


\begin{center}

 {\Large \bf Quantum Scattering Theoretical Description of
Thermodynamical Transport Phenomena}

\vspace{0.5cm} {\bf Tooru Taniguchi}

\vspace{0.1cm} {\it Department of Applied Physics, Tokyo Institute of
Technology, \\ Ohokayama, Megro-ku, Tokyo 152, Japan}

\end{center}


\begin{quote}
\vspace{0.2cm}

\baselineskip 0.5cm

   \hspace{0.6cm} We give a method of describing thermodynamical
transport phenomena, based on a quantum scattering theoretical
approach.
   We consider a quantum system of particles connected to
thermodynamical reservoirs by leads.
   The effects of the reservoirs are imposed as an asymptotic
condition at the end of the leads.
   We derive an expression for a current of a conserved quantity,
which is independent of the details of the Hamiltonian operator.
   The Landauer formula and its generalizations are derived from this
method.

\vspace{0.6cm}
\end{quote}


\baselineskip 0.65cm

   {\bf $ \; \langle$ 1. Introduction $ \; \rangle$ \hspace{0.5cm}}
   Statistical mechanical description of thermodynamical responses
has been one of the important subjects of nonequilibrium statistical
mechanics. Some methods have been proposed for this purpose (for
example, see Ref. [1]).
   Of special interest is a method pioneered by Landauer [2], who
heuristically derived an expression for an electric current,
employing a scattering theoretical approach.
   His method was generalized to some cases; for example, a
multichannel case [3-5], a case of a finite temperature [4,6], a case
of a heat current [6,7] and a case of an inelastic scattering process
caused by a random potential [8].
   These methods describe linear responses of a system to
thermodynamical gaps of reservoirs which induce Fermi distributions
in the system.
   However, these have not explicitly treated an effect of inelastic
scattering processes caused by scatterers, and applications of these
methods have been mainly restricted to mesoscopic phenomena.
   Moreover, there have not existed a unified statistical mechanical
derivation of all the generalizations of Landauer formula [9].

   The purpose of the present Letter is to give a statistical
mechanical method for descriptions of thermodynamical responses,
based on a quantum scattering theoretical approach.
   We show that Landauer formula and its generalizations are derived
by this method.
   This method covers all the cases which have been contained in the
generalizations of Landauer formula.
   Moreover, it can be applied to some new cases; nonlinear responses
to thermodynamical gaps of more general reservoirs inducing non-Fermi
distributions in a system, inelastic processes caused by scatterers,
and currents other than an electric or a heat current, etc.
   So, this method can give new generalizations of Landauer formula.


   {\bf $ \; \langle$ 2. Set-up $ \; \rangle$ \hspace{0.5cm}}
   We consider a quantum system of particles in a three-dimensional
region $\Omega$.
   The system consists of two kinds of particles, which we call
`transport particles' and `scatterers'.
   The transport particles are in a scattering state, and the
scatterers are in a bound state.
   The region $\Omega$ consists of a finite region $\Omega_0$ and $N$
semi-infinite columned regions $\Omega_j, \, j=1,2,\cdots,N$.
   The semi-infinite columned region $\Omega_j$ connects the
region $\Omega_0$ to infinity.
   We call the columned region the `lead'.

   For simplicity, we treat a system of only two particles; one
transport particle and one scatterer.
   We assume the Hamiltonian operator $\hat{H}$ of this system to be
of the form

     \begin{equation}
       \hat{H} \equiv \frac{1}{2m}\left
\{\hat{\Bfp}-\frac{q}{c}\BfA(\hat{\Bfx})\right \}^2 +
\frac{1}{2M}\left \{ \hat{\BfP}-\frac{Q}{c}\BfA(\hat{\BfX})\right
\}^2 + U(\hat{\Bfx}, \hat{\BfX})
     \label{Hamil.1}
     \end{equation}

\noindent where $c$ is the velocity of light, $m$ and $M$ are the
masses of the transport particle and the scatterer, respectively, $q$
and $Q$ are the charges of the transport particle and the scatterer,
respectively, $\hat{\Bfx}$ and $\hat{\BfX}$ are the coordinate
operators of the transport particle and the scatterer, respectively,
$\hat{\Bfp}$ and $\hat{\BfP}$ are the momentum operators of the
transport particle and the scatterer, respectively, $\BfA(\hat{\Bfx})$
and $\BfA(\hat{\BfX})$ are the vector potential operators acting on
the transport particle and the scatterer, respectively, $U(\hat{\Bfx},
\hat{\BfX})$ is the potential operator of the transport particle and
the scatterer.
   Here the square of a vector means the inner product of the vector
with itself.

   The state of this system at time $t$ is described by a density
operator $\hat{\rho}(t)$ which obeys the Liouville-von Neuman equation

     \begin{equation}
       i \hbar \frac{d\hat{\rho}(t)}{dt} = [\hat{H},
            \hat{\rho}(t)]
     \label{Liouv}
     \end{equation}

\noindent where $2\pi\hbar$ is the Planck constant.

   We introduce $\mid \Bfx, \BfX \; \rangle$ as the eigenstate of the
operators $\hat{\Bfx}$ and $\hat{\BfX}$ with eigenvalues $\Bfx$ and
$\BfX$, respectively.
   We introduce the unit vectors ${\Bfe}_k^{\scriptscriptstyle (j)}$,
$k=1,2,3$ as a basis of $\BfR^3$ such that
${\Bfe}_1^{\scriptscriptstyle (j)}$ is parallel to the $j$-th
columned region and is pointing to the finite region $\Omega_0$.
   We define $\hat{x}_k^{\scriptscriptstyle (j)}$ by
$\hat{x}_k^{\scriptscriptstyle (j)} \equiv
{\Bfe}_k^{\scriptscriptstyle (j)} \cdot \hat{\Bfx}$, and introduce
$x_k^{\scriptscriptstyle (j)}$ as a eigenvalue of
$\hat{x}_k^{\scriptscriptstyle (j)}$ ($j=1,2,\cdots,N$, $k=1,2,3$).
   We assume the functions $\BfA(\Bfx)$ and $U(\Bfx, \BfX)$ to have
the asymptotic forms satisfying

     \begin{equation}
       \BfA(\Bfx) \stackrel{x _1^{\scriptscriptstyle
(j)}\rightarrow-\infty}{\sim} \BfA^{\scriptscriptstyle
(j,\infty)}(x_2^{\scriptscriptstyle (j)}, x_3^{\scriptscriptstyle
(j)}), \;\;\;\;\; \mbox{in} \;\;\Bfx \in \Omega_j,
     \end{equation}

     \begin{equation}
       U(\Bfx, \BfX) \stackrel{x_1^{(j)}\rightarrow-\infty}{\sim}
U^{\scriptscriptstyle (j,\infty)}(x_2^{\scriptscriptstyle (j)},
x_3^{\scriptscriptstyle (j)}, \BfX), \;\;\;\;\; \mbox{in} \;\;\Bfx
\in \Omega_j
     \end{equation}

\noindent where $\BfA^{\scriptscriptstyle
(j,\infty)}(x_2^{\scriptscriptstyle (j)}, x_3^{\scriptscriptstyle
(j)})$ is a function of $x_2^{\scriptscriptstyle (j)}$ and
$x_3^{\scriptscriptstyle (j)} $, and $U^{\scriptscriptstyle
(j,\infty)}(x_2^{\scriptscriptstyle (j)}, x_3^{\scriptscriptstyle
(j)}, \BfX)$ is a function of $x_2^{\scriptscriptstyle (j)}$,
$x_3^{\scriptscriptstyle (j)} $ and $\BfX$ only. We consider the
operator $\hat{H}^{\scriptscriptstyle (j,\infty)}$ defined by

    \begin{eqnarray}
      \hat{H}^{\scriptscriptstyle (j,\infty)} \equiv
         && \!\!\!\!\!\!\!\!\!
\frac{1}{2m}\left\{\hat{\Bfp}-\frac{q}{c}\BfA^{\scriptscriptstyle
(j,\infty)}(\hat{x}_2^{\scriptscriptstyle (j)},
\hat{x}_3^{\scriptscriptstyle (j)})\right\}^2 \nonumber \\
         && +
\frac{1}{2M}\left\{\hat{\BfP}-\frac{Q}{c}\BfA(\hat{\BfX})\right\}^2 +
U^{\scriptscriptstyle (j,\infty)}(\hat{x}_2^{\scriptscriptstyle (j)},
\hat{x}_3^{\scriptscriptstyle (j)}, \hat{\BfX})
     \label{Hamil.2}
     \end{eqnarray}

\noindent The eigenstate $\mid \Phi_{k n}^{\scriptscriptstyle
(j,\infty)} \; \rangle$ of the operator $\hat{H}^{\scriptscriptstyle
(j,\infty)}$ can be represented as

     \begin{eqnarray}
       \mid \Phi_{k n}^{\scriptscriptstyle (j,\infty)} \; \rangle =
\; \mid \phi_{k}^{\scriptscriptstyle (j)} \; \rangle \; \otimes \mid
\varphi_{k n}^{\scriptscriptstyle (j,\infty)} \; \rangle.
     \label{eigen.1}
     \end{eqnarray}

\noindent Here, $\mid \phi_{k}^{\scriptscriptstyle (j)} \; \rangle$
is the eigenstate of the operator $\Bfe_1^{\scriptscriptstyle (j)}
\cdot
\hat{\Bfp}$ with the eigenvalue $\hbar k$ and have an orthonormality

     \begin{eqnarray}
       \langle \; \phi_{k}^{\scriptscriptstyle (j)} \; \mid
\phi_{k'}^{\scriptscriptstyle (j)} \; \rangle = 2\pi \delta(k - k').
     \end{eqnarray}

\noindent And $\mid \varphi_{k n}^{\scriptscriptstyle (j,\infty)} \;
\rangle$ is introduced as the eigenstate of the operator
$\hat{\cal{H}}_k^{\scriptscriptstyle (j,\infty)}$ which is defined by
$\langle \; \phi_{k}^{\scriptscriptstyle
(j)}\mid\hat{H}^{\scriptscriptstyle (j,\infty)}\mid
\phi_{k'}^{\scriptscriptstyle (j)} \; \rangle \equiv
\hat{\cal{H}}_k^{\scriptscriptstyle (j,\infty)}\langle \;
\phi_{k}^{\scriptscriptstyle (j)} \; \mid
\phi_{k'}^{\scriptscriptstyle (j)} \; \rangle$.
   We introduce $E_{kn}^{\scriptscriptstyle (j)}$ as the eigenvalue
of the operator $\hat{H}^{\scriptscriptstyle (j,\infty)}$
corresponding to the eigenstate $\mid \Phi_{kn}^{\scriptscriptstyle
(j,\infty)} \; \rangle$.
   And $\mid x_2^{\scriptscriptstyle (j)}, x_3^{\scriptscriptstyle
(j)}, \BfX \; \rangle$ is introduced as the eigenstate of the
operators $\hat{x}_2^{\scriptscriptstyle (j)}$,
$\hat{x}_3^{\scriptscriptstyle (j)}$ and $\hat{\BfX}$ with
eigenvalues $x_2^{\scriptscriptstyle (j)}$, $x_3^{\scriptscriptstyle
(j)}$ and $\BfX$, respectively.
   We assume an orthonormality

      \begin{equation}
        \int_{S_j}dx_2^{\scriptscriptstyle
(j)}dx_3^{\scriptscriptstyle (j)}\int_{\Omega}d\BfX \langle \;
\varphi_{kn}^{\scriptscriptstyle (j,\infty)} \mid
x_2^{\scriptscriptstyle (j)}, x_3^{\scriptscriptstyle (j)}, \BfX \;
\rangle \, \langle \; x_2^{\scriptscriptstyle (j)},
x_3^{\scriptscriptstyle (j)}, \BfX \mid
\varphi_{kn'}^{\scriptscriptstyle (j,\infty)} \; \rangle =
\delta_{nn'}
      \label{ortho.1}
      \end{equation}

\noindent of the states $\{ \mid \varphi_{kn}^{\scriptscriptstyle
(j,\infty)} \; \rangle\}_{n}$, where $S_j$ represents the projection
of the cross section of the $j$-th columned region onto the
$x_2^{\scriptscriptstyle (j)}x_3^{\scriptscriptstyle (j)}$ plane.

   We assume that there exists an eigenstate $\mid
\Psi_{kn}^{\scriptscriptstyle (j)} \; \rangle$ of the Hamiltonian
operator $\hat{H}$ with the eigenvalue $E_{kn}^{\scriptscriptstyle
(j)}$.
   Here, the eigenstate $\mid \Psi_{kn}^{\scriptscriptstyle (j)} \;
\rangle$ satisfies the following asymptotic condition:

      \begin{eqnarray}
        \langle \; \Bfx, \BfX\mid\Psi_{kn}^{\scriptscriptstyle (j)}
\; \rangle \stackrel{\mid\Bfx\mid \rightarrow \infty}{\sim}
\hspace{8.5cm} \nonumber
      \end{eqnarray}
      \begin{eqnarray}
           \hspace{0.5cm} \left\{ \begin{array}{c} \langle \; \Bfx,
\BfX\mid \Phi_{kn}^{\scriptscriptstyle (j,\infty)} \; \rangle \;\; +
{\displaystyle \sumthree{k'n'}{kk'<0}{E_{k'n'}^{\scriptscriptstyle
(j)}=E_{kn}^{\scriptscriptstyle (j)}}} r_{\scriptscriptstyle
(k',n':k,n)}^{\scriptscriptstyle (j)}\langle \; \Bfx,
\BfX\mid\Phi_{k'n'}^{\scriptscriptstyle (j,\infty)} \; \rangle, \\
           \hspace{6cm} \mbox{in} \;\; \Bfx\in\Omega_{j}
\vspace{0.8cm} \\
           {\displaystyle
\sumthree{k'n'}{kk'<0}{E_{k'n'}^{\scriptscriptstyle
(l)}=E_{kn}^{\scriptscriptstyle (j)}}} t_{\scriptscriptstyle
(k',n':k,n)}^{\scriptscriptstyle (l,j)}\langle \; \Bfx,
\BfX\mid\Phi_{k'n'}^{\scriptscriptstyle (l,\infty)} \; \rangle,
\hspace{3cm} \\
           \hspace{7.1cm} \mbox{in} \;\; \Bfx\in\Omega_{l}, \;\; l
\neq j \end{array}\right.
       \label{asymp.3}
       \end{eqnarray}

\noindent where $r_{\scriptscriptstyle
(k',n':k,n)}^{\scriptscriptstyle (j)}$ and $t_{\scriptscriptstyle
(k',n':k,n)}^{\scriptscriptstyle (l,j)}$ are constants determined by
the fact that $\mid \Psi_{kn}^{\scriptscriptstyle (j)} \; \rangle$ is
an eigenstate of the operator $\hat{H}$.
   It should be noted that the eigenstate $\mid
\Psi_{kn}^{\scriptscriptstyle (j)} \; \rangle$ may not be determined
uniquely by the asymptotic condition (\ref{asymp.3}).
   So we should introduce a new suffix in order to distinguish the
eigenstates having the same asymptotic form.
   But in order to avoid a complicated notation we do not write such
a suffix explicitly, but distinguish such degenerate states by
different values of $n$ from now on.

   We assume that the condition

      \begin{equation}
        \lim_{\mid {\small \BfX} \mid \rightarrow \infty} \langle \;
\Bfx, \BfX \mid \hat{\rho}(t) \mid {\cal{X}} \; \rangle = 0
      \label{bound}
      \end{equation}

\noindent is satisfied for an arbitrary state $\mid \cal{X} \;
\rangle$.
   The condition (\ref{bound}) may be satisfied by the assumption
that the scatterer is in a bound state.
   The transport particle is assumed to be in a scattering state
which satisfies the asymptotic condition

      \begin{eqnarray}
        \langle \; \Bfx,\BfX\mid && \!\!\!\!\!\!\!\!\!\!
\hat{\rho}(t)\mid\Bfx',\BfX' \; \rangle \stackrel{\mid\Bfx\mid
\rightarrow\infty, \mid\Bfx'\mid\rightarrow\infty}{\sim} \nonumber \\
        && \sum_{j=1}^{N}\sumtwo{kn}{(k>0)} F_j(k, n) \langle \; \Bfx,
\BfX\mid\Psi_{kn}^{\scriptscriptstyle {(j)}} \; \rangle \langle \;
\Psi_{kn}^{\scriptscriptstyle (j)}\mid\Bfx',\BfX' \; \rangle
      \label{asymp}
      \end{eqnarray}

\noindent where $F_j(k, n)$ is a positive function of $k$ and $n$.
   We interpret that the function $F_j(k, n)$ represents a
distribution induced in the system by an interaction with the
reservoir at the end of the $j$-th lead.


   {\bf $ \; \langle$ 3. Current of a Conserved Quantity $ \;
\rangle$ \hspace{0.5cm}} We consider a conserved quantity.
   This quantity corresponds to a Hermitian operator $\hat{G}$, which
satisfies the relation $[\hat{H}, \hat{G}] = 0$.

   We introduce the symmetrized product
$\hat{\cal{X}}\star\hat{\cal{Y}}$ of two arbitrary operators
$\hat{\cal{X}}$ and $\hat{\cal{Y}}$ as

     \begin{eqnarray}
        \hat{\cal{X}}\star\hat{\cal{Y}} \equiv \{\hat{\cal{X}}
\hat{\cal{Y}}+(\hat{\cal{X}} \hat{\cal{Y}})^{\dagger}\}/2
       \label{product}
     \end{eqnarray}

\noindent where $(\hat{\cal{X}} \hat{\cal{Y}})^{\dagger}$ means
the Hermitian conjugate operator of the operator $\hat{\cal{X}} \hat{\cal{Y}}$.
   By Eqs. (\ref{Liouv}), (\ref{bound}), (\ref{product}) and
Gauss' divergence theorem we obtain an equation of continuity

     \begin{eqnarray}
       \frac{\partial}{\partial t} \int_{\Omega} d\BfX &&
         \!\!\!\!\!\!\!\!\!\! \langle \; \Bfx, \BfX \mid
(\hat{\rho}(t)\star\hat{G}) \mid \Bfx, \BfX \; \rangle \nonumber \\
&&
         + \frac{\partial}{\partial \Bfx} \cdot \int_{\Omega} d\BfX
\langle \; \Bfx, \BfX \mid ((\hat{\rho}(t)\star\hat{G}) \star
\hat{\Bfv}) \mid \Bfx, \BfX \; \rangle = 0
     \label{conti.2}
     \end{eqnarray}

\noindent where  $\hat{\Bfv}$ is defined by $\hat{\Bfv} \equiv
- [\hat{H}, \hat{\Bfx}] /(i\hbar)$.
   By Eq. (\ref{conti.2}) we interpret the quantity $\int_{\Omega}
d\BfX \langle \; \Bfx, \BfX \mid ((\hat{\rho}(t)\star\hat{G}) \star
\hat{\Bfv}) \mid \Bfx, \BfX \; \rangle$ as a current density of
the conserved quantity due to the transport particle at time $t$.

   We consider the current $J_j$ in the $j$-th lead, defined by

     \begin{eqnarray}
       \!\!\! J_{j}\equiv\lim_{x_1^{\scriptscriptstyle
(j)} \rightarrow -\infty}\int_{S_j}dx_2^{\scriptscriptstyle
(j)}dx_3^{\scriptscriptstyle (j)} \; \Bfe_1^{\scriptscriptstyle
(j)}\cdot &&
         \!\!\!\!\!\!\!\!\!\! \int_{\Omega} d\BfX \langle \; \Bfx,
\BfX \mid ((\hat{\rho}(t)\star\hat{G}) \star \hat{\Bfv}) \mid \Bfx,
\BfX \; \rangle.
     \label{flow}
     \end{eqnarray}

\noindent We assume that the eigenstate
$\mid\Psi_{kn}^{\scriptscriptstyle (j)} \; \rangle$ is also the
eigenstate of the operator $\hat{G}$ [14].
   And we introduce $G_{kn}^{\scriptscriptstyle (j)}$ as the
eigenvalue of the operator $\hat{G}$ corresponding to the eigenstate
$\mid\Psi_{kn}^{\scriptscriptstyle (j)} \; \rangle$.
   From Eqs. (\ref{asymp}) and (\ref{flow}) we derive

     \begin{eqnarray}
       J_{j}  && \!\!\!\!\!\!\!\!\!\! =
          \frac{1}{\hbar}\sumtwo{kn}{(k>0)}F_{j}(k, n) \,
G^{\scriptscriptstyle (j)}_{kn}\left\{ \frac{\partial
E_{kn}^{\scriptscriptstyle (j)}}{\partial k}\right.   -
\sumthree{k'n'}{k'>0}{E_{-k'n'}^{\scriptscriptstyle
(j)}=E_{kn}^{(j)}}\mid r_{\scriptscriptstyle (-k', n',:k,
n)}^{\scriptscriptstyle (j)}\mid^2\left.\frac{\partial
E_{-k'n'}^{\scriptscriptstyle (j)}}{\partial k'}\right\}
\hspace{0.3cm} \nonumber \\
           && - \sumtwoo{l=1}{(l\neq
j)}\frac{1}{\hbar}\sumtwo{kn}{(k>0)}F_{l}(k, n) \,
G^{\scriptscriptstyle (l)}_{kn}
\sumthree{k'n'}{k'>0}{E_{-k'n'}^{\scriptscriptstyle
(j)}=E_{kn}^{\scriptscriptstyle (l)}}\mid t_{\scriptscriptstyle (-k',
n':k, n)}^{\scriptscriptstyle (j,l)}\mid^2\frac{\partial
E_{-k'n'}^{\scriptscriptstyle (j)}}{\partial k'}
         \label{resul}
         \end{eqnarray}

\noindent where we used the following fact [15]:

     \begin{eqnarray}
       \mbox{If} && \!\!\! E_{kn}^{\scriptscriptstyle
(j)}=E_{k'n'}^{\scriptscriptstyle (j)} \; \mbox{and} \; kk'>0, \;
\mbox{then} \nonumber \\
       && \int_{S_j}dx_2^{\scriptscriptstyle
(j)}dx_3^{\scriptscriptstyle (j)}\int_{\Omega} d\BfX \nonumber \\
       && \hspace{1cm} \frac{1}{2}\left\{ \langle \;
\Bfx,\BfX\mid(\Bfe_{1}^{\scriptscriptstyle
(j)}\cdot{\hat{\Bfv}})\mid\Phi_{kn}^{\scriptscriptstyle (j,\infty)}
\; \rangle \langle \; \Phi_{k'n'}^{\scriptscriptstyle
(j,\infty)}\mid\Bfx,\BfX \; \rangle\right.  \nonumber \\
        && \hspace{2cm} + \left.\langle \;
\Bfx,\BfX\mid\Phi_{kn}^{\scriptscriptstyle (j,\infty)} \; \rangle
\langle \; \Phi_{k'n'}^{\scriptscriptstyle
(j,\infty)}\mid(\Bfe_{1}^{\scriptscriptstyle
(j)}\cdot{\hat{\Bfv}})\mid\Bfx,\BfX \; \rangle\right\} \nonumber \\
        && \hspace{3.5cm} \stackrel{\mid x_{1}^{\scriptscriptstyle
(j)}\mid\rightarrow\infty}{\sim} \frac{1}{\hbar}\frac{\partial
E_{kn}^{\scriptscriptstyle (j)}}{\partial k} \, \delta_{n' n} \,
\delta_{k' k}, \;\;\;\;\; \mbox{in} \; \Bfx\in\Omega_j.
      \label{ortho.2}
      \end{eqnarray}

\noindent By the equation of continuity (\ref{conti.2}), the current
of the quantity $G$ which flows through any cross section of the
$j$-th columned region takes the value $J_j$ in any steady state.
   It is important to note that the effect of interference among the
incident wave, its reflective waves and its transmitted waves does
not appear in the quantity $J_j$, because of Eq. (\ref{ortho.2}).
   Eq. (\ref{resul}) is one of the main results in the present Letter.


   {\bf $ \; \langle$ 4. Derivation of Landauer Formula and its
Generalizations $ \; \rangle$
   \hspace{0.5cm}} As a special case, we consider a case satisfying
the following conditions:
   (A) The scattered wave has the same wave number as its incident
wave, that is, when $r_{\scriptscriptstyle (-k', n':k,
n)}^{\scriptscriptstyle (j)} \neq 0$, $kk'>0$ and
$E_{-k'n'}^{\scriptscriptstyle (j)}=E_{kn}^{\scriptscriptstyle (j)}$
in $ n'$, $k$, $ n$ on ${\cal F}_{j}$, or $t_{\scriptscriptstyle
(-k', n':k, n)}^{\scriptscriptstyle (l,j)}\neq 0$, $kk'>0$ and
$E_{-k'n'}^{\scriptscriptstyle (l)}=E_{kn}^{\scriptscriptstyle (j)}$
in $k$, $ n$ on ${\cal F}_{j}$ and in $ n'$ on ${\cal F}_{l}$, the
relation $k'= k$ is satisfied, where ${\cal F}_j$ is the support of
the function $F_{j}(k, n)$.
   It implies that the scattering processes are almost elastic.
   (B) The distribution function $F_j(k, n)$ is a function only of
the energy eigenvalue $E_{kn}^{\scriptscriptstyle (j)}$.
   So we put a function $\tilde{F}_j(E_{kn}^{\scriptscriptstyle
(j)})$ of $E_{kn}^{\scriptscriptstyle (j)}$ instead of $F_j(k, n)$.
   (C) The operator $\hat{G}$ is a function only of the operator
$\hat{H}$.
   So we put an operator $\tilde{G} (\hat{H})$, which is a function of
$\hat{H}$, instead of $\hat{G}$.
   (D) The coefficients $ r_{\scriptscriptstyle (k', n':k,
n)}^{\scriptscriptstyle (j)}$ and $ t_{\scriptscriptstyle (k', n':k,
n)}^{\scriptscriptstyle (l,j)}$ are dependent on the suffixes $k$
and $n$ on ${\cal F}_{j}$ only through the energy eigenstate
$E_{kn}^{\scriptscriptstyle (j)}$.
   So we put the coefficients
$ r_{\scriptscriptstyle (k', n')}^{\scriptscriptstyle (j)}
(E_{kn}^{\scriptscriptstyle (j)})$ and $t_{\scriptscriptstyle (k', n')}
^{\scriptscriptstyle (l,j)}(E_{kn}^{\scriptscriptstyle (j)})$ instead
of the coefficients $ r_{\scriptscriptstyle (k', n':k,
n)}^{\scriptscriptstyle (j)}$ and $ t_{\scriptscriptstyle (k', n':k,
n)}^{\scriptscriptstyle (l,j)}$, respectively.
   (E) The function $E_{kn}^{\scriptscriptstyle (j)}$ of
the variable $k$ projects the domain $(0, \infty)$ of the variable
$k$  to the domain $\Lambda_j$ of the
variable $\varepsilon$.
   The domain $\Lambda_j$ is independent of the value of the suffix $n$.
   (F) The suffix $n$ takes $\tilde{n}$ of values on ${\cal F}_j$.
    Under the conditions (A)-(F), Eq. (\ref{resul}) becomes

      \begin{eqnarray}
        J_{j} = \frac{\tilde{n}}{2\pi\hbar} \left\{\int_{\Lambda_j}
\right.  d\varepsilon && \!\!\!\!\!\!\! \tilde{F}_j(\varepsilon) \,
\tilde{G} (\varepsilon)\left ( 1- R_j(\varepsilon) \right ) \nonumber
\\
        && - \sumtwoo{l=1}{(l\neq
j)} \left.\int_{\Lambda_l}d\varepsilon \tilde{F}_{l}(\varepsilon)
\, \tilde{G} (\varepsilon) T_{\scriptscriptstyle jl}(\varepsilon)\right\}.
       \label{resul.2}
       \end{eqnarray}

\noindent Here $R_j(\varepsilon)$ and $T_{\scriptscriptstyle jl}
(\varepsilon)$ are defined by
$R_j(\varepsilon)\equiv\sum_{k'n'}\mid r_{\scriptscriptstyle (-k', n')}
^{\scriptscriptstyle (j)} (\varepsilon) \mid^2 $
and $T_{\scriptscriptstyle jl}(\varepsilon)
\equiv\sum_{k'n'}\mid t_{\scriptscriptstyle (-k', n')}
^{\scriptscriptstyle (j,l)}(\varepsilon) \mid^2$,
respectively where the sums over the suffixes $k'$ and $n'$
are taken only over values satisfying the condition
$E_{-k'n'}^{\scriptscriptstyle (j)}=\varepsilon$ and $E_{-k'n'}
^{\scriptscriptstyle (l)}=\varepsilon$, respectively.
   And we used the fact that the sum over the suffix $k$ in $k>0$
corresponds to the integral over the suffix $k$ in $k>0$ multiplied
the factor $1/(2\pi)$.

   From Eq. (\ref{resul.2}) we can derive Landauer formula and its
generalizations which have already been proposed.
   For example, if $\tilde{G}(\hat{H})=q$,
$\tilde{F}_j(\varepsilon)=\lim_{T\rightarrow
0}\{\exp\{(\varepsilon-\mu_j)/(k_B
T)\}+1\}^{-1}=\theta(\varepsilon-\mu_j)$ (where $k_B$, $T$ and
$\mu_j$ are positive constants), $\mid (\mu_j-\bar{\mu})/
\bar{\mu}\mid<<1$ (where
$\bar{\mu}\equiv(\mu_1+\mu_2+\cdots+\mu_N)/N$), $\tilde{n}=2$ and
$\Lambda_j=(0,\infty)$, then Eq. (\ref{resul.2}) become

     \begin{eqnarray}
       J_{j} \approx \frac{q}{\pi\hbar}\left\{\left ( 1-
R_j(\bar{\mu}) \right )\mu_j\right. + \sumtwoo{l=1}{(l\neq
j)} \left. T_{\scriptscriptstyle jl}(\bar{\mu}) \;
\mu_l\right\} \label{resul.3}
     \end{eqnarray}

\noindent in the first approximation of
$(\mu_j-\bar{\mu})/\bar{\mu}$.
   Eq. (\ref{resul.3}) is one of the generalizations of Landauer
formula, and was proposed by M. B\"{u}ttiker [5].


    {\bf $ \; \langle$ 5. Conclusion and Remarks $ \; \rangle$
    \hspace{0.5cm}} In the present Letter, we have described
thermodynamical nonlinear responses of a quantum system to
thermodynamical gaps of reservoirs.
    Here, the thermodynamical reservoirs were connected to the system
with leads, and effects of the reservoirs were introduced as an
asymptotic condition at the ends of the leads.
    Thermodynamical gaps of the reservoirs cause thermodynamical
transport phenomena.
    We derived an expression of a current of a conserved quantity,
which is independent of the details of the Hamiltonian operator.
    For example, this method can describe an energy current which is
caused inside a system connected to heat reservoirs having different
temperatures.

    This method leads to Landauer formula and its generalizations
which have already been proposed.
    Moreover, we can also give further generalizations of Landauer
formula by this method.
    For example, in this method we can deal with nonlinear responses,
reservoirs inducing non-Fermi distribution in a system, inelastic
processes caused by scatterers and currents other than an electric or
a heat current, etc., which have not been treated earlier.
    The method in the present Letter can be generalized to the case
of many transport particles and many scatterers [16].
    (However the two-particles system, which we discussed in the
present Letter, can be interpreted as a mean field approximation for
a system consisting of many transport particles and many
scatterers.)
    This method can also be generalized to the case where
interactions of the particles are given by a complex potential only
in a finite region.

    We interpreted that the function $F_j(k, n)$ of $k$ and $n$
represents a distribution induced in the system by an interaction
with thermodynamical reservoir at the end of the $j$-th lead, because
it represents a distribution of the incident wave at the end of the
$j$-th lead, and we can use an equilibrium distribution function as
the stationary distribution function $F_j(k, n)$.
    But these reservoirs do not cause a dissipation of the system in
this method.
    And as the distribution function $F_j(k, n)$ we can select a
distribution function which is not an equilibrium distribution
function.
    In this sense, we can describe non-thermodynamical transport
phenomena by this method.

    One may notice that this method does not treat distributions of
thermodynamic quantities inside a system, because we do not assume
the local equilibrium assumption, etc. in this method.
     For example, this method does not treat temperature distribution
inside a system connected to heat reservoirs having different
temperatures.
     To treat such a distribution by a generalization of this method
remains as a future problem.


    {\bf $\langle \; $ Acknowledgements $ \; \rangle$ \hspace{0.5cm}}
    I wish to express my gratitude to K. Kitahara and T. Hara for a
careful reading of the manuscript and for valuable comments.
    I acknowledge S. Komiyama for a stimulating lecture on Landauer
formula.


\vspace{1cm}

\begin{center}
--------------------------------------------------------
\end{center}

\baselineskip 0.55cm

\noindent [1] D. N. Zubarev, Nonequiliblium Statistical Mechanics,
Consultant Bureau, New York (1971)

\noindent [2] R. Landauer, Philos. Mag. 21 (1970) 863

\noindent [3] M. Y. Azbel, J. Phys. C14 (1981) L225

\noindent [4] M. B\"{u}ttiker, Y. Imry, R. Landauer and S. Pinhas,
Phys. Rev.  B31 (1985) 6207

\noindent [5] M. B\"{u}ttiker, Phys. Rev. Lett. 57 (1986) 1761

\noindent [6] H. L. Engquist and P. W. Anderson, Phys. Rev.  B24
(1981) 1151

\noindent [7] U. Sivan and Y. Imry, Phys. Rev.  B33 (1986) 551

\noindent [8] Y. Gefen and G. Sh\"{o}n, Phys. Rev.  B30 (1984) 7323

\noindent[9] There {\it is} a statistical mechanical derivation of a
generalization of Landauer formula for an electric current in a
restricted situation where the system consists of non-interacting
electrons, reservoirs connected to the system do not have temperature
gaps, and the system causes only elastic scattering processes
[10-13].
   But it should be noted that the methods used in Refs. [10-13] are
not same with our method.
   Refs. [10-12] used the linear response theory to an external
disturbance derived from the potential energy, and Ref. [13] used a
different asymptotic condition with our method.

\noindent[10] E. N. Economou and C. M. Soukoulis, Phys. Rev. Lett. 46
(1981) 618

\noindent[11] D. S. Fisher and P. A. Lee, Phys. Rev.  B23 (1981) 6851

\noindent[12] H. U. Baranger and A. D. Stone, Phys. Rev.  B40 (1989)
8169

\noindent[13] S.  Komiyama and H. Hirai, Phys. Rev. B54 (1996) 2067

\noindent[14] The exchangeable operators $\hat{H}$ and $\hat{G}$ have
a common eigenstate, and the state $\mid\Psi_{kn}^{\scriptscriptstyle
(j)}\rangle$ is a eigenstate of the operator $\hat{H}$. But the
common eigenstate of operators $\hat{H}$ and $\hat{G}$ may not have
the asymptotic form (\ref{asymp.3}).
   So, this is a new assumption.

\noindent[15] The outline of the derivation of Eq.  (\ref{ortho.2})
is as follows. We first note that the divergence of the integrated
function of $x_2^{\scriptscriptstyle (j)}$ and
$x_3^{\scriptscriptstyle (j)}$ in the left hand side of Eq.
(\ref{ortho.2}) must be zero when $E_{kn}^{\scriptscriptstyle
(j)}=E_{k'n'}^{\scriptscriptstyle (j)}$.
   So the left hand side of Eq. (\ref{ortho.2}) must be zero except
for the case $k=k'$ in $\mid x_1^{\scriptscriptstyle (j)} \mid
\rightarrow \infty$,
$\Bfx\in\Omega_j$ when $E_{kn}^{\scriptscriptstyle
(j)}=E_{k'n'}^{\scriptscriptstyle (j)}$.  Besides, we have $ \langle
\; \Bfx,\BfX\mid(\Bfe_{1}^{\scriptscriptstyle
(j)}\cdot{\hat{\Bfv}})\mid\Phi_{kn}^{\scriptscriptstyle (j,\infty)}
\; \rangle \sim \langle \; \Bfx,\BfX\mid
\frac{1}{\hbar}(\partial{\hat{\cal{H}}}_{k}^{\scriptscriptstyle
(j,\infty)}  / \partial k)\mid\Phi_{kn}^{\scriptscriptstyle (j,\infty)}
\; \rangle$ as $\mid x_{1}^{\scriptscriptstyle
(j)}\mid\rightarrow\infty$ in $\Bfx\in\Omega_j$.  By these and Eq.
(\ref{ortho.1}), we obtain Eq. (\ref{ortho.2}).

\noindent[16] For such a generalization we must deal with energy
eigenstates of the system consisting of the many transport particles
and many scatterers, which must be required to satisfy the Pauli
principle.


\end{document}